\documentclass[conference]{IEEEtran}
\IEEEoverridecommandlockouts
\usepackage{cite}
\usepackage{amsmath,amssymb,amsfonts}
\usepackage{algorithmic}
\usepackage{graphicx}
\usepackage{textcomp}
\usepackage{xcolor}
\usepackage{times}
\usepackage{latexsym}
\usepackage[utf8]{inputenc}
\usepackage{hyperref}
\usepackage{rotating}
\usepackage{multirow}

\usepackage{multirow}
\usepackage[normalem]{ulem}
\useunder{\uline}{\ul}{}

\def\BibTeX{{\rm B\kern-.05em{\sc i\kern-.025em b}\kern-.08em
    T\kern-.1667em\lower.7ex\hbox{E}\kern-.125emX}}
\begin{document}

\makeatletter
\newcommand{\linebreakand}{%
  \end{@IEEEauthorhalign}
  \hfill\mbox{}\par
  \mbox{}\hfill\begin{@IEEEauthorhalign}
}
\makeatother

\title{LTRC\_IIITH's Submission for FinCausal-2023 Shared Task: Financial Document Causality Detection}

\author{
\IEEEauthorblockN{1\textsuperscript{st} Pavan Baswani}
\IEEEauthorblockA{\textit{Language Technologies Research Center, KCIS}
 \\
\textit{IIIT Hyderabad}\\
India \\
hiranmai.sri@research.iiit.ac.in
}
\and
\IEEEauthorblockN{1\textsuperscript{st} Hiranmai Sri Adibhatla}
\IEEEauthorblockA{\textit{Language Technologies Research Center, KCIS} \\
\textit{IIIT Hyderabad}\\
India \\
pavan.baswani@research.iiit.ac.in
}
\and
\linebreakand
\IEEEauthorblockN{2\textsuperscript{nd} Manish Shrivastava}
\IEEEauthorblockA{\textit{Language Technologies Research Center, KCIS} \\
\textit{IIIT Hyderabad}\\
India \\
m.shrivastava@iiit.ac.in
}
}

\maketitle
\begin{abstract}
In this paper, we present our team’s effort in the FinCausal-2023 shared task: span-based cause and effect extraction from financial documents for English. Traditionally, causality extraction tasks have been approached as span extraction or sequence labeling tasks. In our approach, we transform the causality extraction task into a text-generation task, making it more suitable for Large Language Models (LLMs). The goal is to improve the performance of LLMs in extraction tasks while also mitigating the common problem of hallucinations in LLM-generated content. This is achieved by experimenting with different models and prompts to identify the most suitable prompt for the task. In the shared task, our submission stood in third position with an F1 score of 0.54 and an exact match score of 0.08.
\end{abstract}


\section{Introduction}
The FinCausal 2023 shared task~\cite{el2023proceedings}, hosted within the Financial Narrative Processing Workshop~\cite{el2020proceedings,el2021proceedings,el2022proceedings}, is designed to extract cause-and-effect relationships from financial documents. In this context, both the cause and its corresponding effect are identified as specific spans within the original documents.
Comprehending and identifying causality within financial documents is instrumental in gaining a deeper insight into the financial market. Causality information is frequently expressed explicitly in financial documents using familiar indicators like "due to", "caused by", or "as a result of". Yet, in many instances, causal relationships can be inferred by examining the chronological order of events, in the absence of specific patterns. This is particularly relevant in the financial sector, where financial performance is often reported with implicit causal relationships. 
\\
In this paper, we tackle the information extraction problem using neural network models that are sequence-based, and with the framework of text-generation tasks using  Large Language Models (LLMs). Our approach involves fine-tuning pre-trained language models and prompt engineering LLMs to excel in text span classification and generation. We conducted training to create a span-based causality extraction system by fine-tuning the roberta-large model~\cite{liu2019roberta} yielding an F1 score of 0.49. Our top-performing model was based on prompt-engineered ChatGPT, achieving an F1 score of 0.54 in the FinCausal 2023 challenge. Our codebase is available at \url{https://github.com/pavanbaswani/Fincausal_SharedTask-2023}

\section{Dataset}
The objective of the Financial Document Causality Detection Task is to enhance the capability to elucidate the reasons behind changes in the financial landscape, serving as an important step in creating precise and meaningful summaries of financial narratives. This task aims to assess the events or sequences of events that lead to the modification of a financial object or the occurrence of an event within a specified context. It primarily involves detecting relations between elements, making it a relation detection task, focusing on identifying the causal and consequential elements within a causal sentence or text block. Each segment is expected to contain one causal element and one effect. Instances of causal sentences and spans that illustrate cause and effect relationships are provided in Table~\ref{table:training_examples}. The examples from the training dataset indicate that the document can contain multiple effects with a single cause.
\\
This task encompasses two subtasks, one in English and one in Spanish. In both subtasks, the objective is to distinguish elements in the sentence associated with the cause and those linked to the effect. Our participation was in the English subtask. The dataset was compiled from various 2019 financial news articles provided by Qwam \footnote{\url{https://www.qwamci.com/}}, supplemented by SEC data from the Edgar Database \footnote{\url{https://www.sec.gov/edgar/search-and-access}}. Furthermore, the dataset was expanded by incorporating 500 new segments from FinCausal 2022~\cite{mariko2022financial}. The details about the data statistics are detailed in Table~\ref{table:dataset_stats}.

\begin{table*}[!h]
\caption{Examples from Training dataset}
\begin{center}
\resizebox{\textwidth}{!}{
\begin{tabular}{|l|l|l|}
\hline
\multicolumn{1}{|c|}{\textbf{Text}}                                                                                                                                                                                                                                                                                                            & \multicolumn{1}{c|}{\textbf{Cause}}                                                                                                                            & \multicolumn{1}{c|}{\textbf{Effect}}                                                                                                                                        \\ \hline
\begin{tabular}[c]{@{}l@{}}Things got worse when the Wall came down. GDP fell 20\%\\ between 1988 and 1993. There were suddenly hundreds of\\ thousands of unemployed in a country that, under Communism,\\ had had full employment.\end{tabular}                                                                                              & Things got worse when the Wall came down.                                                                                                                      & GDP fell 20\% between 1988 and 1993.                                                                                                                                        \\ \hline
\begin{tabular}[c]{@{}l@{}}Things got worse when the Wall came down. GDP fell 20\%\\ between 1988 and 1993. There were suddenly hundreds of\\ thousands of unemployed in a country that, under Communism,\\ had had full employment.\end{tabular}                                                                                              & Things got worse when the Wall came down.                                                                                                                      & \begin{tabular}[c]{@{}l@{}}There were suddenly hundreds of thousands\\ of unemployed in a country that, under Communism,\\ had had full employment.\end{tabular}            \\ \hline
\begin{tabular}[c]{@{}l@{}}North American Income Trust PLC (LON:NAIT) declared\\ a dividend on Tuesday, September 24th, Upcoming.Co.Uk reports.\\ Stockholders of record on Thursday, October 3rd will be paid\\ a dividend of GBX 1.70 (\$0.02) per share on Friday, October 25th.\\ This represents a dividend yield of 0.56\%.\end{tabular} & \begin{tabular}[c]{@{}l@{}}North American Income Trust PLC (LON:NAIT)\\ declared a dividend on Tuesday, September 24th,\\ Upcoming.Co.Uk reports.\end{tabular} & \begin{tabular}[c]{@{}l@{}}Stockholders of record on Thursday, October 3rd\\ will be paid a dividend of GBX 1.70 (\$0.02) per share\\ on Friday, October 25th.\end{tabular} \\ \hline
\begin{tabular}[c]{@{}l@{}}North American Income Trust PLC (LON:NAIT) declared\\ a dividend on Tuesday, September 24th, Upcoming.Co.Uk reports.\\ Stockholders of record on Thursday, October 3rd will be paid\\ a dividend of GBX 1.70 (\$0.02) per share on Friday, October 25th.\\ This represents a dividend yield of 0.56\%.\end{tabular} & \begin{tabular}[c]{@{}l@{}}North American Income Trust PLC (LON:NAIT)\\ declared a dividend on Tuesday, September 24th,\\ Upcoming.Co.Uk reports.\end{tabular} & This represents a dividend yield of 0.56\%.                                                                                                                                 \\ \hline
\begin{tabular}[c]{@{}l@{}}Conduent has a consensus target price of \$12.64, suggesting\\ a potential upside of 80.87\%.\end{tabular}                                                                                                                                                                                                          & Conduent has a consensus target price of \$12.64                                                                                                               & a potential upside of 80.87\%.                                                                                                                                              \\ \hline
\end{tabular}
}
\label{table:training_examples}
\end{center}
\end{table*}

\begin{table}[!ht]
\caption{Dataset Statistics}
\begin{center}
\resizebox{.8\columnwidth}{!}{
\begin{tabular}{l|c|c|}
\cline{2-3}
                                                     & \textbf{Train} & \textbf{Test} \\ \hline
\multicolumn{1}{|l|}{\textbf{\# Documents}}          & 2949           & 480          \\ %\hline
\multicolumn{1}{|l|}{\textbf{\# Duplicates}}         & 319            & 4             \\ %\hline
\multicolumn{1}{|l|}{\textbf{Avg doc len}}           & 45.94          & 58.06         \\ %\hline
\multicolumn{1}{|l|}{\textbf{(Min, Max) doc len}}    & (7,  267)      & (9, 511)      \\ %\hline
\multicolumn{1}{|l|}{\textbf{Avg cause len}}         & 16.78          &   --          \\ %\hline
\multicolumn{1}{|l|}{\textbf{(Min, Max) cause len}}  & (1, 153)       &   --          \\ %\hline
\multicolumn{1}{|l|}{\textbf{Avg effect len}}        & 18.6           &   --          \\ %\hline
\multicolumn{1}{|l|}{\textbf{(Min, Max) effect len}} & (1, 155)       &   --          \\ \hline
\end{tabular}
\label{table:dataset_stats}
}
\end{center}
\end{table}

\section{System Description}
We outline model types that we explore and compare popular information extraction models with prompt-based models utilizing Large Language Models (LLMs).

\subsection{Sequence labeling models}
We employ the conventional sequence labeling technique, similar to BERT's token classification method~\cite{devlin-etal-2019-bert}, for the identification of spans. We expand on BERT, roBERTa, and various sequence-based models for sequence labeling. This approach facilitates token-level recognition, ensuring precise localization and classification of cause and effect spans. Recently, parameter-efficient models~\cite{liu2022few} have gained prominence. These models concentrate on updating only a small subset of parameters when adapting a pre-trained model to downstream tasks. A noteworthy example of parameter-efficient tuning is Low-Rank Adaptation (LoRA)~\cite{hu2021lora}, which seeks to reduce the number of trainable parameters through low-rank representations. We fine-tuned our dataset using the token classification method of bert-large and roberta-large models~\cite{liu2019roberta}. LoRA was implemented on the large models to enhance storage and training efficiency. With significantly fewer parameters, LoRA streamlines and optimizes the model, making it a preferred choice.

\subsection{Zero-shot Predictions from LLMs}
In addition to traditional span-based tagging, we harness the power of advanced language models to enable zero-shot predictions. Our approach capitalizes on the expertise of three Large Language Models (LLMs): ChatGPT, LLAMA-2 \footnote{\url{https://huggingface.co/meta-llama/Llama-2-7b-chat-hf}}, and OCRA\_mini\_v3\_7b \footnote{\url{https://huggingface.co/pankajmathur/orca_mini_v3_7b}}. This innovative methodology empowers us to make predictions without relying on predefined rules or domain-specific training data in the financial sector.

By harnessing the capabilities of these language models, our approach excels in deciphering the intricate language and domain-specific nuances embedded within financial documents. When provided with a financial document as input, these models adeptly unravel causal relationships and identify cause-and-effect pairs within the document's content, underscoring their remarkable adaptability and proficiency in the financial domain.

\subsubsection{Prompt Engineering}
The success of our system in identifying cause-and-effect relationships within financial documents hinges on the careful design of prompts. In our approach, we utilize three distinct types of prompts, each tailored to address specific aspects of the task. These prompts play a pivotal role in guiding the behavior of our AI models, facilitating the extraction of meaningful insights from the financial data.
\\\\
\textbf{General Prompt (GenPrompt) with Task Short Description:}\\
The \textbf{GenPrompt} (refer Table~\ref{table:general_prompt}) in our system is a basic prompt, describing the task in short. It serves as an initial point of interaction between LLMs and the financial documents. While it does not provide precise task-specific details, it offers a high-level overview that allows LLMs to establish context and direction for their analysis.\\
This general prompt sets the stage for subsequent interactions and ensures that LLMs have a clear understanding of the goal: \textit{identifying cause-and-effect relationships within the financial document}. By providing a concise task description, it initiates the model's engagement with the document in a coherent manner.\\

\begin{table}[!ht]
\caption{\textbf{GenPrompt:} General Prompt with a short description of the task}
\begin{center}
\resizebox{\columnwidth}{!}{
\begin{tabular}{|l|}
\hline
\begin{tabular}[c]{@{}l@{}}\#\#\# Instruction\\ Task: Identify the cause and effect from the given financial context.\\ \\ Output Format: \{\\     'Cause': \textless{}cause-identified-from-context\textgreater{},\\     'Effect': \textless{}effect-identified-from-context\textgreater\\ \}\\ \\ \#\#\# Context: ```\{\}```\\ \#\#\# Response:\end{tabular} \\ \hline
\end{tabular}
}
\label{table:general_prompt}
\end{center}
\end{table}

\textbf{Task-Specific (TaskPrompt) Details with Constraints on Output Format:}\\
The \textbf{TaskPrompt} (refer Table~\ref{table:task_guided_prompt}) dives deeper into the task, offering task-specific details and imposing constraints on the desired output format. It is a critical component of our approach, as it enables the models to align their responses with the specific requirements of cause-and-effect identification within financial documents.\\
This prompt includes constraints of identifying cause and effect within the context, ensuring that LLMs have a precise understanding of the relationships they need to identify. Moreover, it outlines constraints on the output format, ensuring that the generated responses conform to the expected structure and clarity.

\begin{table}[!ht]
\caption{\textbf{TaskPrompt:} Task Guided prompt with constraints on Response}
\begin{center}
\resizebox{\columnwidth}{!}{
\begin{tabular}{|l|}
\hline
\begin{tabular}[c]{@{}l@{}}\#\#\# Instruction\\ Task: Identify the cause and effect from the given financial context (enclosed within in\\ three backticks ```).\\ \\Constraints: \\1) Do not generate any token out of this context. \\2) Just copy from the context. \\3)Also, the text should match with the context (should be case sensitive).\\ \\Output Format: \{\\     'Cause': \textless{}cause-identified-from-context\textgreater{},\\     'Effect': \textless{}effect-identified-from-context\textgreater\\ \}\\ \\ \#\#\# Context: ```\{\}```\\ \#\#\# Response:\end{tabular} \\ \hline
\end{tabular}
}
\label{table:task_guided_prompt}
\end{center}
\end{table}

\textbf{Chain-of-Thought Prompt (CoTPrompt) with Detailed Task Definition:}
The \textbf{CoTPrompt} is the most comprehensive prompt (refer Table~\ref{table:cot_prompt}), inspired by the Chain-of-Thought (CoT) technique, provides a structured approach to the task of \textit{identifying cause-and-effect relationships in financial documents}. This prompt guides the LLMs to follow the instructions step-by-step and provides detailed task definition.\\
This prompt outlines a systematic approach to cause-and-effect analysis, breaking the task into manageable steps. It provides a comprehensive task definition, a precise definition of cause and effect, and a set of guidelines to process the financial document. This approach leverages the capabilities of LLMs to reason step-by-step, ensuring a thorough exploration of the document for causal relationships \cite{kim2023cot}.

\begin{table}[!ht]
\caption{\textbf{CoTPrompt:} Chain-of-thought Prompt with Detailed Instructions}
\begin{center}
\resizebox{\columnwidth}{!}{
\begin{tabular}{|l|}
\hline
\begin{tabular}[c]{@{}l@{}}\#\#\# Instruction\\ You will be given financial document text in the three backticks ``` with "Context:" as prefix.\\ \\ Your task is to identify the `Cause` and `Effect` from the given financial context. \\ \\ Please make sure you read and understand these instructions carefully. Please keep this\\ document open while reviewing, and refer to it as needed.\\ \\ Cause and Effect Definition:\\ The cause and effect is defined as a relation established between two events, where the first\\ event acts as the cause of the second event and the second event is the effect of the first event.\\ One cause can have several effects. A cause is why an event happens. The effect is an event\\ that happens because of cause. The cause and effect occurs based on the following criteria,\\ where cause has to occur before effect, and whenever the cause occurs the effect has to occur.\\ \\ Cause and Effect Identification Steps:\\ 1) Read the given document carefully and understand it.\\ 2) Refer the `Cause and Effect Definition` section and\\ identify the `Cause` and `Effect` from the document.\\ 3) Make sure that the identified text of cause and\\ effect should be substring of the given financial document.\\ 4) Generate the response in JSON format provided in\\ the `Output Format:` section below.\\ \\ Output format: \{\\     'Cause': \textless{}cause-identified-from-context\textgreater{},\\     'Effect': \textless{}effect-identified-from-context\textgreater\\ \}\\ \\ \#\#\# Context: ```\{\}```\\ \#\#\# Response:\end{tabular} \\ \hline
\end{tabular}
}
\label{table:cot_prompt}
\end{center}
\end{table}

By employing variety of prompts, we harness the adaptability of LLMs to cater to diverse financial documents, extracting valuable insights that facilitate informed decision-making within the financial domain. These prompts form the backbone of our approach, providing the necessary guidance and structure to drive the cause-and-effect identification process. To validate the effectiveness of the GenPrompt, TaskPrompt, and CoTPrompt, we have examined several samples from our training data, inputting these prompts into ChatGPT and tabulating the resulting responses. Table~\ref{table:prompt_results} demonstrates that the comprehensive instructions within the prompts significantly enhance our response generation (Where \textbf{GT} denotes the Ground Truth). Furthermore, it's evident that some of the samples result in an exact match.

\begin{table*}
\caption{Prompt Responses from ChatGPT}
\begin{center}
\resizebox{\textwidth}{!}{
\begin{tabular}{|c|ll|}
\hline
\multirow{2}{*}{\textbf{Example-1}}  & \multicolumn{2}{l|}{\begin{tabular}[c]{@{}l@{}}	\textbf{Text:    }Things got worse when the Wall came down. GDP fell 20\% between 1988 and 1993. There were suddenly hundreds of thousands of unemployed in a country that,\\ under Communism, had had full employment.\end{tabular}}                                                                                                                                                                                                                                                                                                                                                                                                                                                                                                                                                                                                                                                                                                                                                                                                                                                                                                                                                                                                                                                                                                                                                                                                                                           \\ \cline{2-3} 
                                     & \multicolumn{1}{l|}{\begin{tabular}[c]{@{}l@{}}\textbf{GT Cause:}                 Things got worse when the Wall came down.\\ \textbf{GenPrompt Cause:}    The Wall came down.\\ \textbf{TaskPrompt Cause:}   Things got worse when the Wall came down.\\ \textbf{CoTPrompt Cause:}   GDP fell 20\% between 1988 and 1993\end{tabular}}                                                                                                                                                                                                                                                                                                                                                                                                     & \begin{tabular}[c]{@{}l@{}}\textbf{GT Effect:}                 GDP fell 20\% between 1988 and 1993.\\ \textbf{GenPrompt Effect:}    GDP fell 20\% between 1988 and 1993, and there were suddenly hundreds\\ of thousands of unemployed in a country that, under Communism, had had full employment.\\ \textbf{TaskPrompt Effect:}   GDP fell 20\% between 1988 and 1993. There were suddenly hundreds\\ of thousands of unemployed in a country that, under Communism, had had full employment.\\ \textbf{CoTPrompt Effect:}    hundreds of thousands of unemployed in a country that, under Communism,\\ had had full employment\end{tabular}                                                                                                                                                                                                  \\ \hline
\multirow{2}{*}{\textbf{Example-2}}  & \multicolumn{2}{l|}{\begin{tabular}[c]{@{}l@{}}	\textbf{Text:    }Shares of LON ANCR traded up GBX 4.50 ($0.06) during mid-day trading on Tuesday, hitting GBX 138.50 ($1.81). 18,080 shares of the company's stock traded\\ hands, compared to its average volume of 35,341.\end{tabular}}                                                                                                                                                                                                                                                                                                                                                                                                                                                                                                                                                                                                                                                                                                                                                                                                                                                                                                                                                                                                                                                                                                                                                                                                                                     \\ \cline{2-3} 
                                     & \multicolumn{1}{l|}{\begin{tabular}[c]{@{}l@{}}\textbf{GT Cause:}                 Shares of LON ANCR traded up GBX 4.50 (\$0.06)\\ during mid-day trading on Tuesday\\ \textbf{GenPrompt Cause:}    Shares of LON ANCR traded up GBX 4.50 (\$0.06)\\ during mid-day trading on Tuesday, hitting GBX 138.50 (\$1.81)\\ \textbf{TaskPrompt Cause:}   Shares of LON ANCR traded up GBX 4.50 (\$0.06)\\ during mid-day trading on Tuesday, hitting GBX 138.50 (\$1.81)\\ \textbf{CoTPrompt Cause:}    Shares of LON ANCR traded up GBX 4.50 (\$0.06)\\ during mid-day trading on Tuesday, hitting GBX 138.50 (\$1.81)'\end{tabular}}                                                                                                                  & \begin{tabular}[c]{@{}l@{}}\textbf{GT Effect:}                 hitting GBX 138.50 (\$1.81).\\ \textbf{GenPrompt Effect:}    18,080 shares of the company's stock traded hands,\\ compared to its average volume of 35,341.\\ \textbf{TaskPrompt Effect:}   18,080 shares of the company's stock traded hands,\\ compared to its average volume of 35,341.\\ \textbf{CoTPrompt Effect:}   18,080 shares of the company's stock traded hands,\\ compared to its average volume of 35,341\end{tabular}                                                                                                                                                                                                                                                                                                                                             \\ \hline
\multirow{2}{*}{\textbf{Example-3}}  & \multicolumn{2}{l|}{	\textbf{Text:    }We are already paying an extra 12 cents a gallon on our gasoline purchases, and a \$10 surcharge on our vehicle registration, ostensibly for road improvements.}                                                                                                                                                                                                                                                                                                                                                                                                                                                                                                                                                                                                                                                                                                                                                                                                                                                                                                                                                                                                                                                                                                                                                                                                                                                                                                                         \\ \cline{2-3} 
                                     & \multicolumn{1}{l|}{\begin{tabular}[c]{@{}l@{}}\textbf{GT Cause:}                 road improvements.\\ \textbf{GenPrompt Cause:}    Paying an extra 12 cents a gallon on our gasoline\\ purchases, and a \$10 surcharge on our vehicle registration, ostensibly\\ for road improvements.\\ \textbf{TaskPrompt Cause:}   We are already paying an extra 12 cents a gallon on\\ our gasoline purchases, and a \$10 surcharge on our vehicle registration,\\ ostensibly for road improvements.\\ \textbf{CoTPrompt Cause:}    We are already paying an extra 12 cents a gallon\\ on our gasoline purchases\end{tabular}}                                                                                                                         & \begin{tabular}[c]{@{}l@{}}\textbf{GT Effect:}                 We are already paying an extra 12 cents a gallon on our gasoline\\ purchases, and a \$10 surcharge on our vehicle registration\\ \textbf{GenPrompt Effect:}    Increased cost for gasoline and vehicle registration, with the stated\\ purpose being road improvements.\\ \textbf{TaskPrompt Effect:}   \\ \textbf{CoTPrompt Effect:}    a \$10 surcharge on our vehicle registration, ostensibly for road\\ improvements\end{tabular}                                                                                                                                                                                                                                                                                                                                             \\ \hline
\multirow{2}{*}{\textbf{Example-4}}  & \multicolumn{2}{l|}{\begin{tabular}[c]{@{}l@{}}	\textbf{Text:    } So far In 2019 EVERY single month the company has continued to post NEW RECORD REVENUES  On June 1st 2019 the company doubled its NET revenues\\ to over \$40 million with TWO more successful acquisitions in Florida.\end{tabular}}                                                                                                                                                                                                                                                                                                                                                                                                                                                                                                                                                                                                                                                                                                                                                                                                                                                                                                                                                                                                                                                                                                                                                                                                                        \\ \cline{2-3} 
                                     & \multicolumn{1}{l|}{\begin{tabular}[c]{@{}l@{}}\textbf{GT Cause:}                 TWO more successful acquisitions in Florida.\\ \textbf{GenPrompt Cause:}    The company's continued success and new record\\ revenues.\\ \textbf{TaskPrompt Cause:}   So far In 2019 EVERY single month the company\\ has continued to post NEW RECORD REVENUES\\ \textbf{CoTPrompt Cause:}    the company doubled its NET revenues to over \$40\\ million with TWO more successful acquisitions in Florida\end{tabular}}                                                                                                                                                                                                                                 & \begin{tabular}[c]{@{}l@{}}\textbf{GT Effect:}                 So far In 2019 EVERY single month the company has continued to post\\ NEW RECORD REVENUES  On June 1st 2019 the company doubled its NET revenues to\\ over \$40 million\\ \textbf{GenPrompt Effect:}    On June 1st, 2019, the company doubled its net revenues to over \$40\\ million with two more successful acquisitions in Florida.\\ \textbf{TaskPrompt Effect:}   On June 1st 2019 the company doubled its NET revenues to over \$40\\ million with TWO more successful acquisitions in Florida.\\ \textbf{CoTPrompt Effect:}    Every single month the company has continued to post NEW RECORD\\ REVENUES in 2019\end{tabular}                                                                                                                                            \\ \hline
\multirow{2}{*}{\textbf{Example-5}}  & \multicolumn{2}{l|}{\begin{tabular}[c]{@{}l@{}}	\textbf{Text:    }Two grandparents, Joanie and Bill Miller from Buffalo, New York, decided to make Apple their Single-Stock Retirement Plan back in 1997. They put \$16,000 into\\ Apple stock. And since then, they've made \$2 million.\end{tabular}}                                                                                                                                                                                                                                                                                                                                                                                                                                                                                                                                                                                                                                                                                                                                                                                                                                                                                                                                                                                                                                                                                                                                                                                                                           \\ \cline{2-3} 
                                     & \multicolumn{1}{l|}{\begin{tabular}[c]{@{}l@{}}\textbf{GT Cause:}                 They put \$16,000 into Apple stock.\\ \textbf{GenPrompt Cause:}    Joanie and Bill Miller's decision to make Apple their\\ Single-Stock Retirement Plan and their \$16,000 investment into Apple\\ stock in 1997.\\ \textbf{TaskPrompt Cause:}   Two grandparents, Joanie and Bill Miller from Buffalo,\\ New York, decided to make Apple their Single-Stock Retirement Plan back\\ in 1997. They put \$16,000 into Apple stock.\\ \textbf{CoTPrompt Cause:}    Two grandparents, Joanie and Bill Miller from Buffalo,\\ New York, decided to make Apple their Single-Stock Retirement Plan back\\ in 1997. They put \$16,000 into Apple stock.\end{tabular}} & \begin{tabular}[c]{@{}l@{}}\textbf{GT Effect:}                 And since then, they've made \$2 million.\\ \textbf{GenPrompt Effect:}    They've made \$2 million since then.\\ \textbf{TaskPrompt Effect:}   And since then, they've made \$2 million.\\ \textbf{CoTPrompt Effect:}   And since then, they've made \$2 million.\end{tabular}                                                                                                                                                                                                                                                                                                                                                                                                                                                                                                       \\ \hline
\multirow{2}{*}{\textbf{Example-6}}  & \multicolumn{2}{l|}{	\textbf{Text:    }NYSEARCA FNCL traded down $0.49 on Friday, reaching $40.83.}                                                                                                                                                                                                                                                                                                                                                                                                                                                                                                                                                                                                                                                                                                                                                                                                                                                                                                                                                                                                                                                                                                                                                                                                                                                                                                                                                                                                                             \\ \cline{2-3} 
                                     & \multicolumn{1}{l|}{\begin{tabular}[c]{@{}l@{}}\textbf{GT Cause:}                 NYSEARCA FNCL traded down \$0.49 on Friday\\ \textbf{GenPrompt Cause:}    NYSEARCA FNCL traded down \$0.49 on Friday.\\ \textbf{TaskPrompt Cause:}   NYSEARCA FNCL traded down \$0.49 on Friday,\\ \textbf{CoTPrompt Cause:}   NYSEARCA FNCL traded down \$0.49 on Friday\end{tabular}}                                                                                                                                                                                                                                                                                                                                                                       & \begin{tabular}[c]{@{}l@{}}\textbf{GT Effect:}                 reaching \$40.83.\\ \textbf{GenPrompt Effect:}    The price of NYSEARCA FNCL reached \$40.83.\\ \textbf{TaskPrompt Effect:}   reaching \$40.83.\\ \textbf{CoTPrompt Effect:}   reaching \$40.83\end{tabular}                                                                                                                                                                                                                                                                                                                                                                                                                                                                                                                                                                         \\ \hline
\multirow{2}{*}{\textbf{Example-7}}  & \multicolumn{2}{l|}{\begin{tabular}[c]{@{}l@{}}	\textbf{Text:    }Analyst Ratings  This is a breakdown of recent ratings and recommmendations for Auris Medical and Elite Pharmaceuticals, as provided by MarketBeat.com. \\ - - Sell Ratings - Hold Ratings - Buy Ratings - Strong Buy Ratings - Rating Score  - Auris Medical - 0 - 0 - 2 - 0 - 3.00  - Elite Pharmaceuticals - 0 - 0 - 0 - 0 - N/A \\ Auris Medical currently has a consensus price target of \$75.00, indicating a potential upside of 2,383.44\%.\end{tabular}}                                                                                                                                                                                                                                                                                                                                                                                                                                                                                                                                                                                                                                                                                                                                                                                                                                                                                                                                                                                            \\ \cline{2-3} 
                                     & \multicolumn{1}{l|}{\begin{tabular}[c]{@{}l@{}}\textbf{GT Cause:}                 Auris Medical currently has a consensus price\\ target of \$75.00\\ \textbf{GenPrompt Cause:}    The breakdown of recent ratings and\\ recommendations for Auris Medical and Elite Pharmaceuticals.\\ \textbf{TaskPrompt Cause:}   This is a breakdown of recent ratings and\\ recommendations for Auris Medical and Elite Pharmaceuticals,\\ as provided by MarketBeat.com\\ \textbf{CoTPrompt Cause:}    Auris Medical currently has a consensus\\ price target of \$75.00, indicating a potential upside of 2,383.44\%.\end{tabular}}                                                                                                                    & \begin{tabular}[c]{@{}l@{}}\textbf{GT Effect:}                 a potential upside of 2,383.44\%.\\ \textbf{GenPrompt Effect:}    Auris Medical currently has a consensus price target of \$75.00,\\ indicating a potential upside of 2,383.44\%.\\ \textbf{TaskPrompt Effect:}   Auris Medical currently has a consensus price target of \$75.00,\\ indicating a potential upside of 2,383.44\%.\\ \textbf{CoTPrompt Effect:}    Analyst Ratings - This is a breakdown of recent ratings and\\ recommendations for Auris Medical and Elite Pharmaceuticals, as provided by\\ MarketBeat.com. - - Sell Ratings - Hold Ratings - Buy Ratings - Strong Buy Ratings -\\ Rating Score - Auris Medical - 0 - 0 - 2 - 0 - 3.00 - Elite Pharmaceuticals -\\ 0 - 0 - 0 - 0 - N/A\end{tabular}                                                               \\ \hline
\multirow{2}{*}{\textbf{Example-8}}  & \multicolumn{2}{l|}{	\textbf{Text:    }The company reported ($0.80) earnings per share (EPS) for the quarter, beating the Zacks' consensus estimate of ($0.87) by \$0.07.}                                                                                                                                                                                                                                                                                                                                                                                                                                                                                                                                                                                                                                                                                                                                                                                                                                                                                                                                                                                                                                                                                                                                                                                                                                                                                                                                                      \\ \cline{2-3} 
                                     & \multicolumn{1}{l|}{\begin{tabular}[c]{@{}l@{}}\textbf{GT Cause:}                 The company reported (\$0.80) earnings per share\\ (EPS) for the quarter\\ \textbf{GenPrompt Cause:}    The company's earnings per share (EPS) for the\\ quarter being reported.\\ \textbf{TaskPrompt Cause:}   The company reported (\$0.80) earnings per share\\ (EPS) for the quarter,\\ \textbf{CoTPrompt Cause:}    The company reported (\$0.80) earnings per share\\ (EPS) for the quarter\end{tabular}}                                                                                                                                                                                                                                             & \begin{tabular}[c]{@{}l@{}}\textbf{GT Effect:}                 beating the Zacks' consensus estimate of (\$0.87) by \$0.07.\\ \textbf{GenPrompt Effect:}    The company beat the Zacks' consensus estimate of (\$0.87) by \$0.07.\\ \textbf{TaskPrompt Effect:}   beating the Zacks' consensus estimate of (\$0.87) by \$0.07.\\ \textbf{CoTPrompt Effect:}    beating the Zacks' consensus estimate of (\$0.87) by \$0.07\end{tabular}                                                                                                                                                                                                                                                                                                                                                                                                                 \\ \hline
\multirow{2}{*}{\textbf{Example-9}}  & \multicolumn{2}{l|}{	\textbf{Text:    }GM stock is down 2.8\% since the UAW strike began, worse than the 0.5\% drop of the Dow Jones Industrial Average over the same span.}                                                                                                                                                                                                                                                                                                                                                                                                                                                                                                                                                                                                                                                                                                                                                                                                                                                                                                                                                                                                                                                                                                                                                                                                                                                                                                                                                    \\ \cline{2-3} 
                                     & \multicolumn{1}{l|}{\begin{tabular}[c]{@{}l@{}}\textbf{GT Cause:}                 the UAW strike began, worse than the 0.5\% drop of the\\ Dow Jones Industrial Average over the same span.\\ \textbf{GenPrompt Cause:}    The UAW strike began.\\ \textbf{TaskPrompt Cause:}   GM stock is down 2.8\% since the UAW strike began,\\ \textbf{CoTPrompt Cause:}   GM stock is down 2.8\% since the UAW strike began\end{tabular}}                                                                                                                                                                                                                                                                                                            & \begin{tabular}[c]{@{}l@{}}\textbf{GT Effect:}                 GM stock is down 2.8\%\\ \textbf{GenPrompt Effect:}    GM stock is down 2.8\% since the strike began, which is worse than\\ the 0.5\% drop of the Dow Jones Industrial Average over the same span.\\ \textbf{TaskPrompt Effect:}   worse than the 0.5\% drop of the Dow Jones Industrial Average\\ over the same span.\\ \textbf{CoTPrompt Effect:}    worse than the 0.5\% drop of the Dow Jones Industrial Average\\ over the same span\end{tabular}                                                                                                                                                                                                                                                                                                                           \\ \hline
\multirow{2}{*}{\textbf{Example-10}} & \multicolumn{2}{l|}{\begin{tabular}[c]{@{}l@{}}	\textbf{Text:    }By 2017, the share of exports was down to 9 percent. This is on par with the share in the United States but is far lower than the shares in Germany (34 percent),\\ South Korea (28 percent), and Japan (14 percent). This shift has been largely obscured because the country's output, imports, and exports have all been rising so dramatically\\ in absolute terms.\end{tabular}}                                                                                                                                                                                                                                                                                                                                                                                                                                                                                                                                                                                                                                                                                                                                                                                                                                                                                                                                                                                                                                                                       \\ \cline{2-3} 
                                     & \multicolumn{1}{l|}{\begin{tabular}[c]{@{}l@{}}\textbf{GT Cause:}                 This shift has been largely obscured because the country's\\ output, imports, and exports have all been rising so dramatically in absolute terms.\\ \textbf{GenPrompt Cause:}    The share of exports in the country decreased to 9 percent\\ by 2017.\\ \textbf{TaskPrompt Cause:}   By 2017, the share of exports was down to 9 percent.\\ \textbf{CoTPrompt Cause:}   By 2017, the share of exports was down to 9 percent\end{tabular}}                                                                                                                                                                                                                & \begin{tabular}[c]{@{}l@{}}\textbf{GT Effect:}                 By 2017, the share of exports was down to 9 percent.\\ \textbf{GenPrompt Effect:}    This share is now on par with the share in the United States but\\ significantly lower than the shares in Germany (34 percent), South Korea (28 percent),\\ and Japan (14 percent). This shift has been obscured by the significant rise in the country's\\ output, imports, and exports in absolute terms.\\ \textbf{TaskPrompt Effect:}   This shift has been largely obscured because the country's output, imports,\\ and exports have all been rising so dramatically in absolute terms.\\ \textbf{CoTPrompt Effect:}    This shift has been largely obscured because the country's output, imports,\\ and exports have all been rising so dramatically in absolute terms\end{tabular} \\ \hline
\end{tabular}
}
\end{center}
\label{table:prompt_results}
\end{table*}

\section{Experiments \& Results}
Initially, we fine-tuned the pretrained transformer models, BERT and roBERTa, for span-based classification to address the FinCausal problem, treating it as a Named Entity Recognition (NER) task.
We partitioned the training dataset into two segments, the training and development sets, with an 80-20 split ratio. Our training utilized a sequence length of 512 tokens and the Adam optimizer with a learning rate of 0.001. The training was conducted on a system with the following specifications: GPU Name - Nvidia P100, GPU Memory - 16GB, GPU Clock - 1.32GHz, CPU Cores - 2, RAM - 12GB, and the platform used was Kaggle.

Simultaneously, we explored responses generated by instruction-tuned models (\textit{ChatGPT, llama-2, and ocra\_mini\_v3\_7b}) using \textit{GenPrompt, TaskPrompt, and CoTPrompt} (detailed in Table~\ref{table:general_prompt}, \ref{table:task_guided_prompt} and \ref{table:cot_prompt} respectively). Our analysis, as depicted in Table~\ref{table:prompt_results}, revealed that the \textit{CoTPrompt} was the most effective prompt for identifying cause-and-effect relationships within financial documents. Subsequently, we adopted this prompt for use with the other models. Table~\ref{table:test_data_results}, shows the results of the models with the exact match metric. Notably, the \textit{ChatGPT} model, when paired with the \textit{CoTPrompt}, outperformed other models, achieving an exact match score of 0.75 in identifying causal relationships.

% Initially, we experiment by fine-tuning pre-trained transformers BERT and roBERTa models on span-based classification (as NER problem). The train set divided into two splits (the training and development set) with 80, 20 ratio respectively. We employed 512 sequence length, Adam optimizer with learning rate 0.001. The specifications used for training are GPU Name: Nvidia P100, GPU Memory: 16GB, GPU Clock: 1.32GHz, CPU Cores: 2, RAM: 12GB, PLATFORM: Kaggle.

% From Table~\ref{table:prompt_results}, we identified that the \textit{CoTPrompt} is the most suitable prompt for identifying the Cause and Effect for the given financial document. Hence, we use this prompt for the other models. 
%Table~\ref{table:test_data_results}, provides the model results with exact match. Compared to other model \textbf{ChatGPT + CoTPrompt} model outperformed in identifying the Cause and Effect with Exact Match as 0.75.

\begin{table*}[!h]
\caption{Experimental Results on Test Dataset}
\begin{center}
\resizebox{0.6\textwidth}{!}{
\begin{tabular}{c|c|c|c|c|}
\cline{2-5}
                                                                                                          & \textbf{Precision} & \textbf{Recall} & \textbf{F1}    & \textbf{Exact Match} \\ \hline
\multicolumn{1}{|c|}{\textbf{BERT-large}}                                                                 & 0.496              & 0.324           & 0.392          & 0.012                \\ \hline
\multicolumn{1}{|c|}{\textbf{roBERTa-large}}                                                              & 0.596              & 0.448           & 0.493          & 0.004                \\ \hline
\multicolumn{1}{|c|}{\textbf{\begin{tabular}[c]{@{}c@{}}ChatGPT\\ +\\ TaskPrompt\end{tabular}}}           & 0.637              & 0.315           & 0.339          & 0.000                \\ \hline
\multicolumn{1}{|c|}{\textbf{\begin{tabular}[c]{@{}c@{}}llama-2\\ +\\ CoTPrompt\end{tabular}}}            & 0.580              & 0.275           & 0.285          & 0.000                \\ \hline
\multicolumn{1}{|c|}{\textbf{\begin{tabular}[c]{@{}c@{}}ocra\_mini\_v3\_7b\\ +\\ CoTPrompt\end{tabular}}} & 0.585              & 0.404           & 0.436          & 0.010                \\ \hline
\multicolumn{1}{|c|}{\textbf{\begin{tabular}[c]{@{}c@{}}ChatGPT\\ +\\ CoTPrompt\end{tabular}}}            & 0.582              & 0.521           & \textbf{0.542} & \textbf{0.075}       \\ \hline
\end{tabular}
}
\label{table:test_data_results}
\end{center}
\end{table*}

Our best-performing model (\textbf{ChatGPT + CoTPrompt}), secured a notable position in this shared task (refer Table~\ref{table:comparitive_results}). It achieved the third position in the exact match metric and the fourth position in the F1 score metric, underscoring its competitive performance in identifying causal relationships within financial documents. When considering the overall ranking, our model solidly clinched the third position, showcasing its effectiveness in this challenging FinCausal problem.

\begin{table*}[!ht]
\caption{Comparitive results over Other submissions of shared task}
\begin{center}
\resizebox{0.6\textwidth}{!}{
\begin{tabular}{|l|c|c|c|c|}
\hline
\multicolumn{1}{|c|}{\textbf{Submission}} & \textbf{Precision} & \textbf{Recall} & \textbf{F1}   & \textbf{Exact Match} \\ \hline
MB                                        & 0.72               & 0.71            & 0.71          & 0.25                 \\ %\hline
Luis                                      & 0.66               & 0.60            & 0.62          & 0.17                 \\ %\hline
\textbf{Ours}                             & 0.58               & 0.52            & \textbf{0.54} & \textbf{0.08}        \\ %\hline
mirunaz                                   & 0.47               & 0.45            & 0.46          & 0.06                 \\ %\hline
neelesh310                                & 0.62               & 0.56            & 0.58          & 0.00                 \\ \hline
\end{tabular}
}
\label{table:comparitive_results}
\end{center}
\end{table*}

\section{Ablation Study}
Accuracy is assessed by comparing the generated strings to the gold standard string using an exact match criterion. Upon scrutinizing the discrepancies in the generated outputs, we consistently identify two predominant error categories.

\subsection{Text Overflow}
"Text Overflow" is a consistent phenomenon across all models based on Large Language Models (LLMs). This condition generates additional text that may or may not be part of the actual document but is not relevant to the cause or effect span. As evident in examples 2, 4, and 10 in Table~\ref{table:prompt_results}, the predicted text spans for causes and effects contain more information than the ground truth. However, the length of the overflow text has notably diminished in the TaskPrompt and CoTPrompt models compared to the GenPrompt model. Remarkably, the CoTPrompt achieved exact matches for examples 6 and 8. This suggests that the integration of few-shot learning and prompt tuning on the cause-effect dataset may lead to improved exact match spans.

\subsection{Cause and Effect swapped}
A recurrent error found in all three prompts is the inadvertent swapping of cause and effect, clearly exemplified in instances 1, 9, and 10 of Table~\ref{table:prompt_results}. This issue arises because the original text only implicitly mentions the cause and effect, making it challenging for the models to accurately identify and classify them. Notably, the CoTPrompt, which includes explicit cause and effect definitions, exhibits an improved comprehension of these relationships. Additionally, it boasts greater robustness compared to other prompts by eliminating blanks in the generated outputs.
\\
Both the above-mentioned errors may be mitigated through prompt tuning and the inclusion of a few examples of cause and effect in a few-shot learning context.

\section{Conclusions \& Future work}
Our primary focus lies in the generation of cause-and-effect span embeddings, achieved through the thoughtful engineering of prompts for Large Language Models (LLMs). Notably, when we applied the Chain-of-thought prompt (CoTPrompt) to ChatGPT, it outperformed other supervised sequence labeling models. Moving forward, we have aspirations to delve deeper into this line of research, exploring techniques such as few-shot learning and prompt tuning on LLMs like llama-2 and ocra\_mini\_v3\_7b.

This approach will hold great promise in leveraging the strengths of both LLMs and supervised models through a combination strategy. When fine-tuned with cause-effect specific data, LLMs should exhibit remarkable aptitude in recognizing and extracting precise cause-and-effect spans, surpassing the zero-shot and few-shot capabilities of LLMs.

\end{document}